# Secure Vehicle Communications Using Proof-of-Nonce Blockchain

N. Y. Ahn [1], D.H. Lee [2]

[1] *Institute of Cyber Security & Privacy at Korea University, Seoul, South Korea*
[2] *Institute of Cyber Security & Privacy &*
*The Graduate School of Information Security at Korea University, Seoul, South Korea*
[1] `humble@korea.ac.kr`

*Abstract*—**This paper presents an autonomous driving that achieves physical layer security. Proposed vehicle communication is implemented based on Proof-of-Nonce (PoN) blockchain algorithm. PoN blockchain algorithm is a consensus algorithm that can be implemented in light weight. We propose a more secure vehicle communication scheme while achieving physical layer security by defecting PoN algorithm and secrecy capacity. By generating a block only when secrecy capacity is greater than or equal to the reference value, traffic information can be provided only to vehicles with physical layer security. This vehicle communication scheme can secure sufficient safety even from hackers based on quantum computing.**

## I. Introduction

Autonomous driving is a collection of cutting-edge technologies and will become the flower of the future industry [1], [2]. The days to use autonomous driving more safely and more conveniently will not be too far. Autonomous driving is being performed based on vehicle-to-vehicle communication [3-5]. The commercialization of autonomous driving is expected to take place in the near future. A fatal weakness in safety in autonomous driving for safe driving has been revealed in recent studies. It is the emergence of the technology of quantum computing [6], [7]. According to recent reports, the commercialization of quantum computing is expected to proceed faster than the use of autonomous driving [8], [9].

There is a significant possibility that the safety of autonomous driving will be destroyed by a person who has acquired quantum computing technology by a malicious user. Ahn and Lee proposed to introduce physical layer security for autonomous driving to solve these problems [7]. They propose to create a security cluster only for vehicles that have secrecy capacity, which forms the basis of physical layer security, and to communicate with vehicles within the security cluster. In addition, recent studies on vehicle communication using blockchain are being introduced [6], [10], [11]. Our paper proposes to form a more secure security cluster by combining PoN blockchain and physical layer security, and to perform vehicle communication within the security cluster. The contribution of our thesis is to achieve autonomous driving by forming a lightweight security cluster and securing physical layer security unlike the existing blockchain-based vehicle communication. In Chapter Ⅱ, we introduce the general concept of PoN blockchain. Chapter III refers to autonomous driving using PoN blockchain. Chapter IV introduces block verification using secrecy capacity for physical layer security. Chapter V compares the advantages and disadvantages of the existing blockchain-based autonomous driving and the proposed PoN-based blockchain-based autonomous driving.

## II. PoN Blockchain for IoT

PoN blockchain uses a distributed consensus algorithm by selecting the subject of the untraceable distributed consensus [12], [13]. PoN Blockchain is known to overcome the limitations of scalability and security of the existing blockchain Proof-of-Work (PoW) and Proof-of-Stake (PoS) methods. Proof-of-Nonce (PoN) distributed consensus algorithm minimizes resource consumption by participating in block generation only among a part of all nodes, makes selection prediction for a block generation authority obtaining node impossible, and at the same time, selects a node that represents the whole probability, In order to prevent the occurrence of Fork by securing selected nodes above the determined value, PoN distributed consensus algorithm cannot arbitrarily damage the book after block confirmation, and can be applied to public and private blockchains.

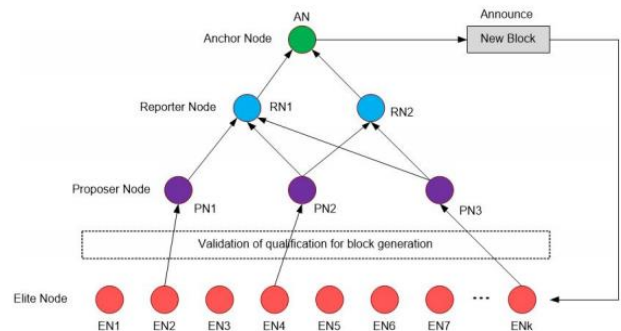

Fig. 1. PoN consensus algorithm

### A. PoN distributed Consensus algorithm

PoN distributed consensus algorithm satisfies the following two conditions. In the first condition, a node having the authority to create a block cannot be predicted in advance. Blocks created under the second condition cannot be modified again. First, the first condition is a part necessary to prevent an attack by collusion between nodes. In the case of public block chains, which no one can trust, there is a high possibility of attack if the node to create the block is known in advance. Therefore, it is necessary to prevent this. In addition, the selected node must be able to confirm its qualification. The second condition is that the agreed block cannot be modified.

The method of randomly selecting a node participating in block generation in PoN distributed consensus algorithm is as follows. One value is calculated by hashing the hash value of the previous block and an arbitrary value generated for each node. By comparing the calculated value with an arbitrary reference value, a node for creating a block can be selected. At this time, all the participating nodes can check whether the node that has generated the block has properly generated the hashed result. PoN algorithm is an algorithm that guarantees the fairness of consensus subjects and provides decentralization. In other words, PoN algorithm provides an opportunity to generate a block that is probabilistically fair to all participants using a verifiable nonce. PoN algorithm provides fairness, transparency, stability, and liveness for decentralization.

Referring to FIG. 1, the nodes constituting PoN consensus algorithm are as follows: an elite node, a proposer node, a reporter node, and an anchor node. Elite nodes are nodes that participate in PoN consensus algorithm. A proposer node is a node that has acquired consensus subject qualification (i.e., candidate block generation authority) for every height among elite nodes. The reporter node is a node that verifies the candidate block generated from the proposer node. The reporter node verifies the qualification of the proposer node or verifies the candidate block. Anchor node (i.e. RSU) is a node that selects the final block from among the verified candidate blocks and updates the selected block with a new block in the blockchain.

To this end, every node knows an arbitrary value at which the block creation participating node has occurred. However, two problems may occur when a node generating a block publishes an arbitrary value of itself. First, an attack by collusion between nodes is possible by pre-calculating a node to participate in block generation based on the published random value. Second, a malicious node can use an arbitrary value of a specific node when it knows an arbitrary value used by a node generating a block. In order to solve these problems, other nodes must be able to confirm that any value used by the node belongs to the corresponding node.

*B. Candidate Node Selection*

Also, nodes participating in the creation of a new block should not be predictable by other nodes in advance. Therefore, a nonce chain is used for the Proof of Nonce (PoN) distributed consensus algorithm. Here, the nonce chain means nonces that are continuously generated from a master key. The master key is a random number that each node keeps private. Base n is a private random number generated using the master key. The nonce chain can be composed of the generated nonce chain and start height to start using the nonce chain by performing a hash of base n repeatedly m times. For example, the last hash value (hash(0)) of the generated nonce chain and the height (start_height) of the block to which the corresponding nonce chain will be used are disclosed to all elite nodes. Thereafter, whether to participate in the node can be determined by fetching a value corresponding to (current height-start height of the block) from the nonce chain and hashing the hash value of the height1 block. For example, suppose that a node created a new nonce chain, and the block height was used from number 10. When selecting nodes to participate in block creation in order to newly create the height block 12, the height of the new block is 12. The start height of the corresponding node is 10, and the difference between the height of the new block and the start height is 2.

Therefore, hash(2) is extracted from the nonce chain owned by the node. The extracted hash(2) is stored for validity of node participation. Subsequently, by publishing the hash(2) used by the other node as the block creation node, it can be proved that he has justly participated in the block creation. In addition, in order to confirm the validity of the node generating the block, the other node repeats the hash(2) hash operation (block height-start height) published by the block generating node, whose value is the block generating node. It can be confirmed that it is the same as the hash(0) value previously disclosed. PoN blockchain can determine the candidate node selection for block creation as follows. Each of the participating nodes can digitally sign the previous block hash value with the private key for each participating node. Here, the previous block The hash value hd may be a head hash value of a previous block.

$$\text{Pre\_blk\_sign} = \text{sign}(\text{Key\_private})(\text{pre\_hash}) \quad (1)$$

By making the previous block hash value (pre_hash) an electronic signature (Pre_blk_sign) for each node with the private key (Key_private) of the participating node, other nodes cannot predict whether to obtain the block creation authority. Each of the nodes cannot predict the digital signature result because the private key (Key_private) of the other node is unknown. In an embodiment, the node that has obtained the authority to generate a candidate block can verify the authority acquisition fact by presenting the digital signature together. In an embodiment, there may be a procedure in which nodes register in advance in a node pool in order to be a block generation candidate.

Also, when registering a node in the pool, it can be used as a private blockchain when the node's registration conditions are strongly controlled. On the other hand, if the registration possibility is open to all nodes participating in the blockchain, it can be used in the public blockchain. The result of the digital signature (Pre_blk_sign) for each participating node can be processed to have uniform distribution characteristics. For example, by hashing the digital signature (Pre_blck_sign) as follows, a value of significance (key; nonce) can be generated.

$$\text{key} = \text{hash}(\text{Pre\_blk\_sign}) \quad (2)$$

Here, the key value has a uniform distribution characteristic by the hash characteristic. In addition, in order to have the above key value, it can be implemented through a hash value of a previous block hash value and a value obtained from a nonce chain.

The above-described candidate node selection method minimizes resource consumption by participating in block generation of only a part of all nodes, makes prediction for a node for obtaining block generation authority impossible, and makes the selected node represent all of the guaranteed probability, It is possible to secure selected nodes over a certain number of stably.

The algorithm operation of PoN distributed agreement may proceed as follows. First candidate blocks may be generated according to a participation node selection method. When one of the generated first candidate blocks is locked (locked), the selected node can operate as the master node of the first block. In an embodiment, when the first master node is selected, the master node of the first block (N-1 Block) informs the number of nodes generating the first candidate block and the distance provided by each node in the second block voting. It can be used as a parameter for voting to be used. In an embodiment, the hash value of the block generated by the master node of the first block may be used as a previous block hash value for generating second candidate blocks. At this time, through the operation described in the method for determining the right to select a candidate node for block generation, each node can calculate whether to obtain the authority of the second block generating node by comparing the coupon and the reference value.

A node that has obtained the second candidate block generation authority may generate a candidate block and distribute it to all nodes. Nodes that generated the first candidate block may participate in voting by passing one or more distances included in the block transmitted by the second candidate nodes received by each node to the first master node. The first master node is the minimum distance (or maximum, distance) agreed upon by a certain ratio (e.g., a majority, 2/3, etc.) of nodes participating in the first block generation in the voting results of the nodes generating the first candidate block. The lock block can be delivered to all nodes in order to lock (generate the final block) of the second candidate block, which is proposed by extraction. At this time, when the second block (N Block) is locked, the block chain can be completed by repeating the above described processes for the third block.

PoN distributed consensus algorithm (PoN) may provide a consensus structure in which Fork does not occur by determining the second master node according to the voting result of the first candidate blocks. In addition, PoN distributed consensus algorithm (PoN) is impossible to damage the book by locking the block so that voting cannot proceed again to change it after the master node is selected. In addition, since PoN distributed consensus algorithm (PoN) determines nodes participating in block updates at every block generation step, it is impossible to predict and attack them. In addition, PoN distributed consensus algorithm (PoN) may improve disadvantages of nodes having many resources, such as PoS, monopolizing update authority by allowing multiple block generating nodes to select multiple opinions voted. In addition, since the process of locking the second block is determined through the first nodes in PoN distributed consensus algorithm (PoN), transaction irregularities such as double payment can be prevented. In addition, PoN distributed consensus algorithm (PoN) can be used in common for both private and public blockchains by managing only nodes participating in candidate block generation.

III. AUTONOMOUS DRIVING USING PoN BLOCKCHAIN

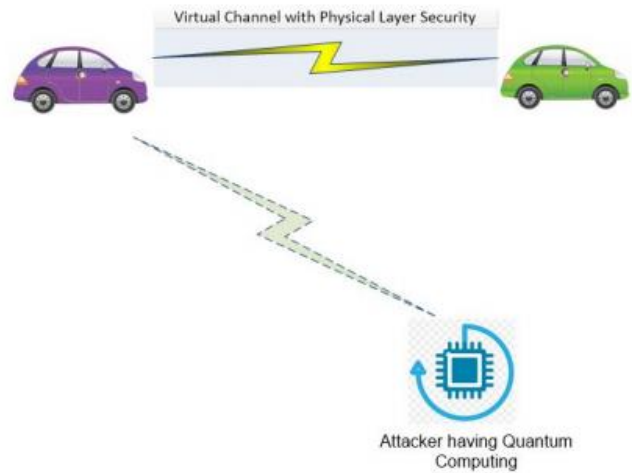

Fig. 2. Autonomous driving using a virtual channel with physical layer security

Autonomous driving is a great benefit to users. This is because you can go to the destination you want to safely and comfortably without any special effort for driving. Autonomous driving basically operates based on V2V communication. The biggest advantage of V2V communication known to us is safe driving. In other words, unlike the situation of driving by a person, it is to prevent traffic accidents using V2V communication. Is it true? Ironically, a bad friend appears in V2V communication for our safe driving. The friend's name is a hacker with quantum computing skills. Oh My God! It is unpredictable what will happen. Since quantum computing technology is already aimed at commercialization, the appearance of hackers with such technology is also expected. It is clear that V2V communication will not guarantee safety by such hackers. It was V2V communication for safety, but now it is in a situation where it cannot guarantee safety.

Fortunately, these issues and related studies are pouring. Ophthalmologist proposed a method using physical layer security using secrecy capacity to perform secure V2V communication from hackers with quantum computing ability. Their proposal is expected to be a realistic and economical answer, helping to bring V2V commercialization. Referring to Fig. 2, physical layer security is simply to be understood as a technology that makes a radiated radio channel into a virtual wired channel. A technique in which vehicles having a certain secrecy capacity form a cluster and perform V2V communication within the formed cluster has also been recently proposed.

## A. Secure Cluster

In this study, we intend to propose a method for forming a secure cluster through blockchain. Research on V2V communication using blockchain is ongoing. However, the existing blockchain is a technology that simply applies the blockchain to V2V communication, not considering the need for physical layer security. No matter how much blockchain is applied, it is almost certain that V2V communication will still face difficulties by hackers with quantum computing technology that is on the verge of commercialization. I believe the need for our research has been fully discussed here.

Although we briefly summarize our research, it is to create a secure cluster by creating and releasing blocks of the blockchain only in vehicles that achieve physical layer security, and perform communication within the formed secure cluster. We used PoN blockchain to form a secure cluster. PoN blockchain is known as a blockchain technology that has increased security with light weight compared to existing blockchains.

## B. Traffic Information

Traffic information for autonomous driving based on PoN blockchain is shown. Here, the traffic information is part of the transaction information of the block. The traffic information may include location information (e.g., GPS), vehicle speed, driving direction information, and time information for each vehicle V1 to Vk. Such traffic information can be used as transaction information of the blockchain block. Vehicles that have secured more than a certain amount of secrecy capacity can create new blocks from the previous blocks. At this time, the generated block may include updated traffic information. For example, the new block may update new traffic information for the second vehicle V2, as shown in figure. However, the traffic information included in the blockchain block we propose will not be limited to this. The traffic information may include link and speed information processed with time and location information of each vehicle. Here, the link is a basic unit in which traffic information providing route information to a vehicle is classified.

## IV. BLOCK VERIFICATION USING SECRECY CAPACITY

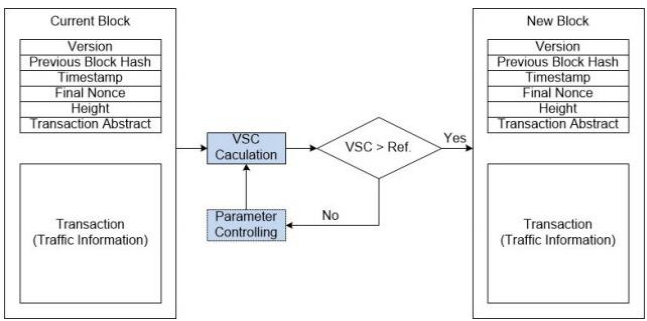

Fig. 3. A proposed scheme for gerating block

Proposed block generation scheme is conceptually shown. Using secrecy capacity, it is determined whether generated block becomes a new block. Each vehicle can create a temporary block with new traffic information. However, creating a temporary block does not make it a block of the blockchain. The blockchain that we propose can be created as a block of the blockchain only for blocks created in vehicles that have passed the secrecy capacity verification. Therefore, each vehicle should basically perform secrecy capacity verification after creating a temporary block.

Fig. 3 is shown that a proposed scheme of generating block is shown in detail. When the security capacity exceeds the reference value, a new block may be generated. And when the security capacity is less than the reference value, the security capacity value can be increased by controlling the security capacity parameter.

If the secrecy capacity is not higher than the reference value, the temporary block will be discarded soon. Thereafter, the corresponding vehicle can be adjusted to a reference value or higher by controlling the secrecy capacity. The above diagram conceptually explains the process of updating a block in our proposed blockchain.

The proposed block can be divided into header information and transaction. The header information may basically include version information, previous block hash value, time stamp, last nonce value, height, transaction summary, and the like. Transactions may include traffic information. The version information may include software/protocol version information. This block hash value means the block hash value of the block immediately preceding the block chain. The final nonce value and height value of the excitation can be used when the block is verified by the opponent. A detailed explanation will be given when explaining PoN blockchain.

In the current block, when attempting to add update information to a blockchain as a new block in a transaction, the secrecy capacity for the vehicle can be calculated. Vehicle secrecy capacity can be calculated in a variety of ways. Ophthalmology was also defined using only SNR values. The vehicle secrecy capacity may use a typical average secrecy capacity. Regardless of how it is defined, the secrecy capacity must be greater than or equal to the reference value to create blocks. When the secrecy capacity is below the reference value, the vehicle secrecy capacity can be made above the reference value by adjusting parameters affecting the secrecy capacity. Ophthalmologist studied various parameters that control the secrecy capacity. According to their claim, the parameters affecting the secrecy capacity may largely include vehicle operation related parameters, antenna related parameters, path related parameters, and noise related parameters. These parameters support the ability to control the secrecy capacity in real time while autonomous driving.

If the vehicle secrecy capacity is greater than or equal to the reference value, the new block can be updated with a new block on the blockchain by propagating to other vehicles. Also, will the blocks made like this be reliable? How can you ensure that reliability? The answer to this can be obtained from PoN blockchain technology. PoN blockchain will be described below.

A hash chain is formed using the number of the vehicle as a unique value, and the hash chain is stored in the vehicle for block generation and in an anchor furnace (for example, RSU) for verifying the generated block. PoN blockchain is using hash chain generation. By publishing the hash chain's final and height values, the final and height values are used to verify the block. In the vehicle blockchain proposed by us, a hash chain is generated using the vehicle's unique number as a nonce value. These hash chains are stored in each vehicle. At the same time as the traffic information, the final and height values of the hash chains are disclosed, so that blocks can be verified in other vehicles.

The vehicle electronic device is implemented to generate a block for autonomous driving, calculate a security capacity, and control the security capacity. The generated block may be transmitted to the outside through an antenna. The ECU for blockchain-based autonomous driving may include a block generator, a secrecy capacity calculator, a discriminator, and a secrecy capacity controller. The block generator may generate blocks corresponding to traffic information. The new block may include a BSM (Basis Short Message). The secrecy capacity calculator can calculate the secrecy capacity. There are various ways to calculate the secrecy capacity.

The discriminator can compare the secrecy capacity and the reference value. As a result of comparison of the discriminator, if the secrecy capacity is larger than the reference value, a new block can be transmitted to the outside. To be transmitted to the outside means that it is combined with the blockchain as a new block. The new propagation block contains real-time traffic information.

## V. COMPARISON OF PoW, PoS AND PoN

| Algorithm | Energy Consumption | Confirmed Time for Consensus |
|---|---|---|
| PoW | Difficulty of Hash Puzzle | $z \cdot t$ |
| PoS | Difficulty of Hash Puzzle | $z \cdot t$ |
| PoN | Secrecy Capacity Control | $T_b + T_q + T_v + T_s$ |

Fig. 4. Performance Comparison of PoW, PoS, and PoN algorithms

In the PoN series algorithm, all nodes have the authority to generate blocks, and blocks are generated by competition, and the probability of block generation success is determined according to computational power. In PoN/PoW algorithms, the energy consumption required to generate a block corresponds to the hash computation time to find the correct answer to the hash puzzle [14]. On the other hand, in PoN algorithm, energy is consumed as much as the time required to control secrecy capacity above a certain level in the electronic device of the vehicle, not the time corresponding to this hash calculation puzzle.

Regardless of the difficulty value of PoW/PoN, this energy consumption may vary depending on how the security capacity parameters are adjusted in the PoN algorithm. Moreover, in vehicles that are considered to have sufficient power compared to other IoT devices, energy consumption will not be a big issue. In fact, the really important issue is consensus confirmation time. In PoN/PoW algorithms, confirmed time is $z \cdot t$. Here z is number of blocks to wait until consensus is reached and t is time of generating the block [15]. In the PoN algorithm, the consensus node time will be determined by the sum of the time $T_b$ to generate one block, the time $T_q$ to verify the block generation qualification, the time $T_v$ to verify the block, and the time $T_s$ to select the final block, referring to Fig. 4. Here, each of $T_q$, $T_v$, and $T_s$ is a very small time compared to the block generation time. In terms of time to reach consensus, it can be seen that the PoN algorithm has very good performance compared to the PoW/PoS algorithms.

## VI. CONCLUSION

We proposed the formation of a security cluster in autonomous driving based on PoN blockchain. PoN blockchain is a lightweight blockchain technology, ensuring safety with low power. In addition, we were not simply grafted to the PoN blockchain, but by verifying only the blocks created in vehicles that expanded secrecy capacity above the reference value as a new block, we were able to achieve physical layer security. The achievement of this physical layer security enabled vehicle communication that secured safety from the potential danger of quantum computing-based hackers in the future. Our research is expected to be of great help in achieving security more safely and efficiently in the era of autonomous driving. In the future, our research is expected to shed even more light in combination with edge computing technology.

ACKNOWLEDGMENT

Thanks to Dr. Oh from ERTI in Korea for introducing PoN blockchain.